\documentclass[12pt]{iopart}
\usepackage{iopams}
\expandafter\let\csname equation*\endcsname\relax

\expandafter\let\csname endequation*\endcsname\relax

\usepackage{bm,bbm}
\usepackage{amsmath}
\usepackage{amsfonts}
\usepackage{amstext,amsthm}
\usepackage{amssymb}   
\usepackage{graphicx}

\begin{document}

\hyphenation{ge-ne-ra-li-zed}
\def\hf{\hat{f}}
\def\Ord{\mathcal{O}}
\newcommand{\barr}{\begin{eqnarray}}
\newcommand{\earr}{\end{eqnarray}}
\newcommand{\beq}{\begin{equation}}
\newcommand{\eeq}{\end{equation}}
\newcommand{\be}{\begin{equation}}
\newcommand{\ee}{\end{equation}}
\newcommand{\ra}{\right\rangle}
\newcommand{\de}{\mathrm{d}}
\newcommand{\la}{\left\langle}
\newcommand{\correc}{ \textcolor{red} }
\newtheorem{theorem}{Theorem}

\newcommand{\gv}[1]{\ensuremath{\mbox{\boldmath$ #1 $}}} 
\newcommand{\uv}[1]{\ensuremath{\mathbf{\hat{#1}}}} 
\newcommand{\abs}[1]{\left| #1 \right|} 
\let\underdot=\d 
\renewcommand{\d}[2]{\frac{d #1}{d #2}} 
\newcommand{\dd}[2]{\frac{d^2 #1}{d #2^2}} 
\newcommand{\pd}[2]{\frac{\partial #1}{\partial #2}} 
\newcommand{\pdd}[2]{\frac{\partial^2 #1}{\partial #2^2}} 
\newcommand{\pdc}[3]{\left( \frac{\partial #1}{\partial #2}
 \right)_{#3}} 
\newcommand{\ket}[1]{\left| #1 \right>} 
\newcommand{\bra}[1]{\left< #1 \right|} 
\newcommand{\braket}[2]{\left< #1 \vphantom{#2} \right|
 \left. #2 \vphantom{#1} \right>} 
\newcommand{\matrixel}[3]{\left< #1 \vphantom{#2#3} \right|
 #2 \left| #3 \vphantom{#1#2} \right>} 
\newcommand{\grad}[1]{\gv{\nabla} #1} 
\let\divsymb=\div 
\renewcommand{\div}[1]{\gv{\nabla} \cdot #1} 
\newcommand{\curl}[1]{\gv{\nabla} \times #1} 
\let\baraccent=\= 
\renewcommand{\=}[1]{\stackrel{#1}{=}} 

\newcommand{\numberset}{\mathbb}
\newcommand{\N}{\numberset{N}}
\newcommand{\Z}{\numberset{Z}}
\newcommand{\R}{\numberset{R}}
\newcommand{\C}{\numberset{C}}

\newcommand{\Res}{\mathrm{Res}}
\newcommand{\Var}{\mathrm{Var}}
\newcommand{\Cov}{\mathrm{Cov}}
\newcommand{\avg}[1]{\left< #1 \right>} 
\newcommand{\T}{\mathcal{T}}
\newcommand{\rs}{\rho^{\star}}
\def\im{{\rm i}}
\newcommand{\kk}{\kappa}
\newcommand{\EE}{\mathcal{E}}
\newcommand{\Cc}{\mathcal{C}}

\def\lm{\lambda_-}
\def\lp{\lambda_+}
\def\lpm{\lambda_\pm}

\def\Wt{\tau_\mathrm{W}}
\def\Ht{\tau_\mathrm{H}}
\def\Sm{\mathcal{S}}

\def\Ht{\tau_\mathrm{H}}
\def\Nc{N}
\def\Sm{\mathcal{S}}
\def\rho{\varrho}

\title[Universality of the weak pushed-to-pulled transition ]{Universality of the weak pushed-to-pulled transition in systems with repulsive interactions}

\author{Fabio Deelan Cunden$^{1}$, Paolo Facchi$^{2,3}$, Marilena Ligab\`o$^{4}$ and Pierpaolo Vivo$^{5}$}

\address{                 
$1.$ School of Mathematics and Statistics, University College Dublin, Belfield, Dublin 4, Ireland\\
$2.$  Dipartimento di Fisica and MECENAS, Universit\`a di Bari, I-70126 Bari, Italy\\
$3.$ Istituto Nazionale di Fisica Nucleare (INFN), Sezione di Bari, I-70126 Bari, Italy\\
$4.$ Dipartimento di Matematica, Universit\`a di Bari, I-70125 Bari, Italy\\
$5.$  King's College London, Department of Mathematics, Strand, London WC2R 2LS, United Kingdom
}

\begin{abstract}
We consider a $d$-dimensional gas in canonical equilibrium under  pairwise screened Coulomb repulsion and external confinement, and subject to a volume constraint (hard walls). We show that its excess free energy displays a  third-order singularity separating the pushed and pulled phases, irrespective of range of the pairwise interaction, dimension and details of the confining potential. The explicit expression of the excess free energy is universal  and interpolates between the Coulomb (long-range) and the delta (zero-range) interaction. The order parameter of the transition---the electrostatic \emph{pressure} generated by the surface excess charge---is  determined by invoking a fundamental energy conservation argument.
\end{abstract}

\maketitle

\section{Introduction} The understanding of when and how \emph{phase transitions} occur---i.e. instances whereby certain properties of a medium change, often abruptly, as a result of the change of some external condition, such as temperature, pressure, or others---is one of the most striking successes of classical statistical mechanics. Formally, phase transitions arise when a thermodynamic potential like the  \emph{free energy} $F$  displays non-analytic point(s) as a function of one driving parameter, and can be classified according to the regularity of $F$ at the transition point. First-order and second-order (discontinuous first or second derivative, respectively) transitions are amongst the most common textbook examples. Recently, a class of weaker transitions (third-order) of the \emph{pushed-to-pulled} type has attracted much attention and has been investigated in a wealth of physics problems related at various levels to random matrices~\cite{Allez14,Atkin14,Borot12,Cunden15,Cunden16a,Cunden16,Dean06,Dean08,DePasquale10,Facchi08,Gross80,Johansson98,LeDoussal16,Majumdar14,Nadal11,VMB08,Vivo10,Wadia80,Wolchover14}. 
\par
The general setting can be formulated as follows. Consider a classical $d$-dimensional gas of $N$ particles in canonical equilibrium at inverse temperature $\beta$, with positional energy
\begin{equation}
E(x_1,\dots,x_N)=\frac{1}{2}\sum_{i\neq j}\Phi(x_i-x_j)+N\sum_i V(x_i),\qquad (x_i\in\R^d).\label{energy1}
\end{equation}
Here, $\Phi(x)$ is a pairwise repulsion kernel, while $V(x)$ is a confining potential. 
\par
In the absence of further constraints, as $N\to\infty$---which is simultaneously a thermodynamic and zero-temperature limit---the particles will arrange in an equilibrium configuration under the  competing mutual repulsion $\Phi(x_i-x_j)$ and global confinement $V(x_i)$. The prefactor $N$ in the external potential ensures that, for large $N$, both terms in the energy are of same order $\Or(N^2)$, with the particles confined in a region of order $\Or(1)$. 
When $N\to\infty$, the equilibrium configuration can be characterized by the \empty{density} of the gas, which, under general assumptions~\cite{Spohn99,Chafai14}, is the minimizer of the mean-field free energy functional at zero temperature 
\be
\EE[\rho]=\frac{1}{2}\iint\Phi(x-y)\rho(x)\rho(y)\de x\de y+\int V(x)\rho(x)\de x,
\label{eq:EF1}
\ee
where the entropic term is absent. 
The quantity $N^2 \EE[\rho]$ 
is the `continuum version' of the energy~\eqref{energy1}. 
Hereafter, the densities $\rho(x)\geq0$ are normalized to $1$. 
\begin{figure}[t]
\centering
\includegraphics[width=.65\columnwidth]{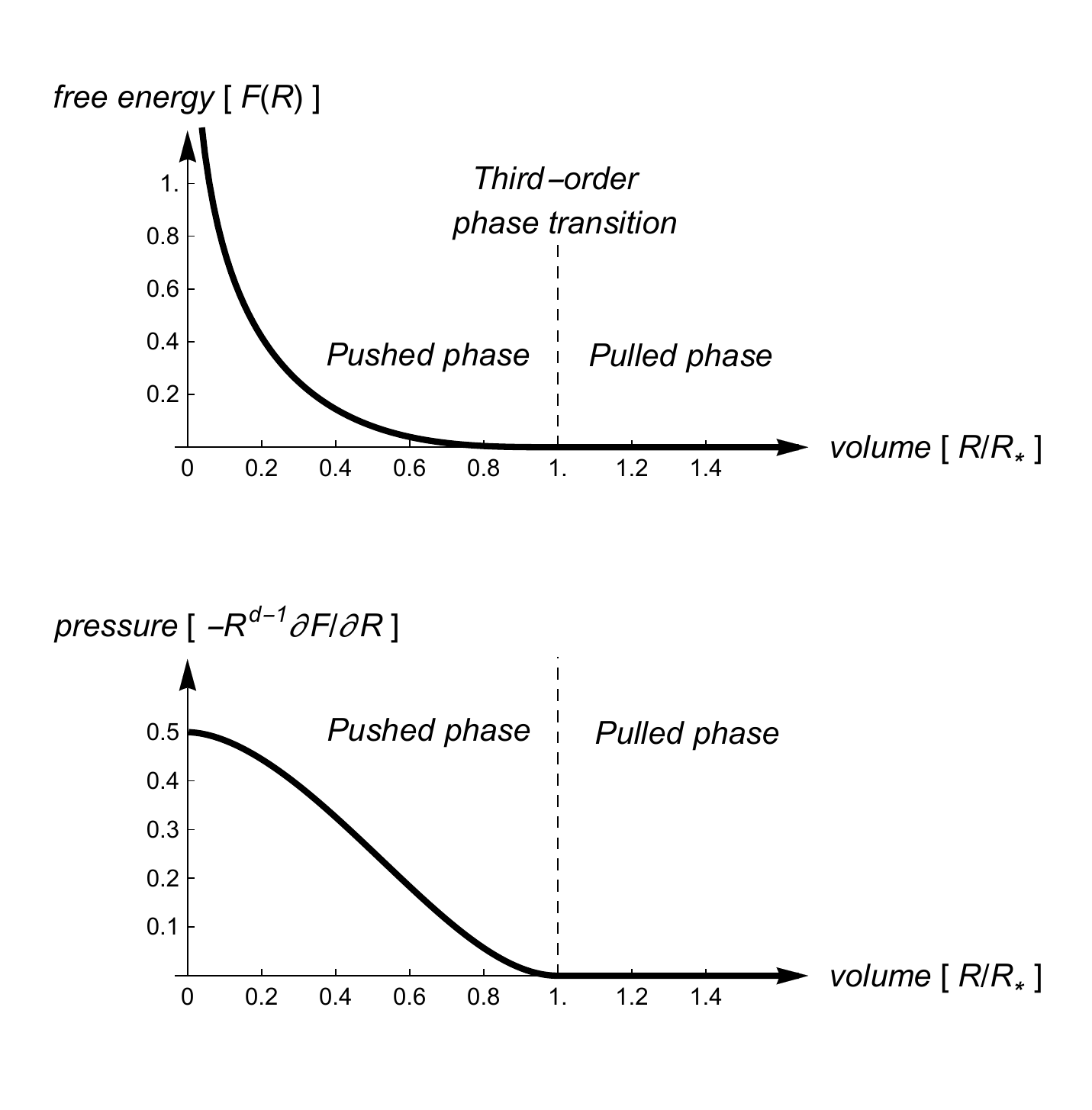}
\caption{TOP: In the pulled phase ($R>R_{\star}$) the volume constraint is immaterial. In the pushed phase ($R<R_{\star}$) the free energy of the gas increases as the gas gets more and more confined. At the transition $R=R_{\star}$ the free energy has a third-order singularity. BOTTOM: Order parameter of the transition: the pressure (in rescaled units) due to the excess charge on the surface of the gas. Here $d=2$, $a=m=1$ and $v(x)=x^2/2$ (see main text).
}
\label{fig:transition}
\end{figure}
\par
What does the minimizer of~\eqref{eq:EF1} look like, then? For isotropic kernels $\Phi(x)=\varphi(|x|)$ and radial confining potentials $V(x)=v(|x|)$, the equilibrium density  inherits the radial symmetry and is supported on a ball $B_{R_\star}$ of radius $R_{\star}>0$. For example, the two-dimensional Coulomb gas, $\varphi(r)=- \log r$, confined by the quadratic potential $v(r)=r^2/2$ fills uniformly the unit disk ($R_\star=1$) in the complex plane, a fact that is known as the \emph{circular law} for  the Ginibre ensemble~\cite{Ginibre65} of non-Hermitian random matrices.
\par
What happens now if we instead force the gas within a smaller ball $B_R$ (of radius $R<R_\star$) than it would normally occupy if unconstrained? The gas will `readjust itself' in a new equilibrium configuration with density $\rho_R(x)$ supported on $B_R\subset B_{R_{\star}}$, and its free energy will increase. 
\par
Denoting by $F(R)=\EE[\rho_R]-\EE[\rho_{R_\star}]$ the \emph{excess free energy} (see Fig.~\ref{fig:transition}), there is overwhelming evidence that generically
\be
F(R)\simeq (R_{\star}-R)^3, \quad\text{as }R\uparrow R_{\star},
\label{eq:3rdorder1}
\ee
implying that the transition between the \emph{pushed} and \emph{pulled} phases of the gas is third-order (see Fig.~\ref{fig:transition}).
\par
When $\Phi(x)=-\log|x|$ (\emph{log-gas}) and $d=1$ or $d=2$, this singularity has been established for quadratic confinements~\cite{Allez14,Atkin14,Cunden16,Dean06,Dean08} (GUE and GinUE ensembles of random matrices and their $\beta>0$ generalisations).  
Up until recently,  singularities in the free energies due to volume constraints have not been systematically investigated in systems with repulsive interactions other than logarithmic. 
In a previous work~\cite{Cunden17}, we have proved that~\eref{eq:3rdorder1} holds true if $\Phi(x)$ is the $d$-dimensional Coulomb interaction for all $d\geq1$, and any convex and smooth potential $V(x)$. 
\par
The ubiquity of this transition calls for a comprehensive theoretical framework, which should be valid irrespective of spatial dimension $d$ and the details of the confining potential $V$, and for the widest class of repulsive interactions $\Phi$. In this Letter, we provide a unified theory for the class of local interaction kernels $\Phi=\Phi_d$ satisfying
\be
\mathrm{D}\Phi_d(x)=\Omega_d\delta(x),\quad\text{with}\quad
\mathrm{D}=-a^2\Delta+m^2, 
\label{eq:Yukawa_eq1vv}
\ee
where $\Omega_d=2\pi^{d/2}/\Gamma\left(d/2\right)$ is the surface area of the unit sphere $S^{d-1}$ ($\Omega_1=2$, $\Omega_2=2\pi$, $\Omega_3=4\pi$, etc.) and $a,m\geq 0$.  
The reason why this kernel is especially relevant is twofold: on one hand,~\eqref{eq:Yukawa_eq1vv} naturally interpolates between the Coulomb electrostatic potential in free space (long-range, for $a=1$ and $m=0$), and the delta-like interaction (short-range, for $a=0$ and $m=1$), while intermediate values $a,m>0$ correspond to the Yukawa (or screened Coulomb) potential. On the other hand, the formulation of the problem in such general terms allows us to identify the previously elusive
order parameter---a quantity that vanishes in one phase (pulled) but is nonzero in the other (pushed)---of this third-order transition: the `electrostatic' \emph{pressure} generated by the surface excess charge. 
\par
So far, the commonly accepted ground for a weak pushed-to-pulled transition to occur has been the existence of \emph{long-range} interactions among the gas particles, this being also one of the main distinctive features of each and every random matrix-related instances of it. We show here instead that this ingredient is not at all needed.
\par
The first task is to compute the constrained equilibrium density $\rho_R(x)$ corresponding to the interaction kernel $\Phi_d$, which is the minimizer of the quadratic functional~\eqref{eq:EF1}.
\par
\section{Constrained equilibrium measure} A variational argument
 determines the necessary Euler-Lagrange (E-L) conditions for $\rho_R(x)$ to be a minimizer in the ball $B_R$ (see, e.g.,~\cite{Bernoff11})
\be
\begin{cases}
\displaystyle\int\Phi_d(x-y)\rho_R(y)\de y=\mu(R)-V(x)&\text{a.e. in $\operatorname{supp}\rho_R$},\\
\displaystyle\int\Phi_d(x-y)\rho_R(y)\de y\geq\mu(R)-V(x)&|x|\leq R,
\end{cases}
\label{eq:E-L}
\ee
where the chemical potential $\mu(R)$ is a constant fixed by the normalization condition $\int_{x|\leq R}\rho_R(x)\de x=1$. Note that in the pushed phase $\operatorname{supp}\rho_R=B_R$. Physically, Eqs.~\eqref{eq:E-L} guarantee that the energy density 
\be
e(x)= \int\Phi_d(x-y)\rho_R(y)\de y +V(x)
\ee
is constant (and equals $\mu(R)$) within the support, and is larger outside it, so that moving any portion of charge outside the support of $\rho_R(x)$ is bound to increase the total electrostatic energy. 
\par
How can the E-L conditions be used in practice? If the interaction kernel satisfies Eq.~\eqref{eq:Yukawa_eq1vv}, then applying the operator $\mathrm{D}$ to both sides of the first condition~\eqref{eq:E-L} yields the equation
$
\rho_R(x)=\mathrm{D}\left[\mu(R)-V(x)\right]/\Omega_d$
for a.e.\ $x$ in the support. 
Once the constrained equilibrium measure $\rho_R(x)$ is known,   its excess free energy $F(R)=\EE[\rho_R]-\EE[\rho_{R_\star}]$ can be computed.
\par
A couple of remarks are still in order. First, if the interaction kernel has strictly positive Fourier transform $\widehat{\Phi}_d(k)>0$, as in our case, then the minimization problem above has a unique solution. 
Second, one can prove the absence, at equilibrium, of condensation of particles \emph{within the bulk}, i.e. absence of $\delta$-components. Condensation of particles, though not possible in the bulk, may however occur on the boundary of the support, and this phenomenon will be a crucial ingredient in the development of the theory.  
\par
We are now ready to apply this general formalism first to the case of long-range interactions (\emph{Coulomb gas}, $\mathrm{D}=-\Delta$) for a radial confinement $V(x)=v(|x|)$, already discussed in~\cite{Cunden17} and included here to prepare the ground for a unified theory.
 \par
\section{Coulomb interaction} Let $\Phi_d(x)$ be the Coulomb electrostatic potential in free space, i.e., the solution of
$
-\Delta\Phi_d(x)=\Omega_d\delta(x)$ for $x\in \mathbb{R}^d$ ($d\geq 1$),
which can be written as $\Phi_d(x)=\varphi_d(|x|)$, where $\varphi_d(r)=(d-2)^{-1}r^{2-d}$ if $d\neq 2$ and $\varphi_d(r)=-\log r$ if $d=2$.
\par
In the unconstrained problem, the equilibrium density of the Coulomb gas is supported on the ball of radius $R_{\star}$, defined as the smallest positive solution of
$
R_\star^{d-1}v'(R_\star)=1
$.
In the pushed phase, instead, the equilibrium density in the bulk does not change, while the excess charge accumulates on the surface. 
Hence, the minimizer of the constrained problem is~\cite{Cunden17}
\be
\rho_R(x)=\frac{1}{\Omega_d}\Big[\Delta V(x)\, \mathbbm{1}_{|x|\leq  R\wedge R_{\star}}
+ \frac{c(R)}{R^{d-1}}\, \delta(R-|x|) \Big],
\label{eq:equilibriummeas}
\ee
where 
\be
R\wedge R_{\star} = \min\{R, R_{\star}\},
\ee 
and the excess charge $c(R)$ is fixed by the normalization 
$
c(R)=\left(1-R^{d-1} v'(R)\right)\mathbbm{1}_{R\leq R_{\star}}
$. See Fig.~\ref{fig:mp}. Note that $c(R_\star)=0$, as it should. Eq.~\eqref{eq:equilibriummeas} expresses the well-known fact that, at the electrostatic equilibrium, any excess charge is distributed on the surface of a conductor, i.e. the boundary of  $\operatorname{supp}\rho_R$~\cite{Jackson}.  
\par
A direct calculation
yields the excess free energy~\cite[Eq.~(31)]{Cunden17}, which can be interestingly cast in the appealing form
\be
F(R)=\frac{1}{2} \int_{R\wedge R_{\star}}^{R_{\star}}  \frac{c(r)^2}{r^{d-1}}\de r,
\label{eq:rateFgeneral}
\ee
a quantity that directly involves the \emph{square of the excess charge} $c(r)$. From~\eqref{eq:rateFgeneral}, since $c(R_{\star})=0$, one shows that indeed 
$F(R)\sim c'(R_{\star})^2 (R_{\star}-R)^3 / 6 R_{\star}^{d-1}$, as  $R\uparrow  R_{\star}$.
\par
The above findings  prompt two important questions. First, would the third-order singularity survive if the range of the pairwise interaction were much shorter? And in that case, would the expression~\eqref{eq:rateFgeneral} still hold?
\par
To investigate these issues, we now turn to a zero-range model in the same class as Eq.~\eqref{eq:Yukawa_eq1vv} for $a=0$, which has the advantage of being exactly solvable too.
\par
\section{Delta-interaction} 
\label{sec:delta}
\begin{figure*}[ht]
\centering
\includegraphics[width=1\columnwidth]{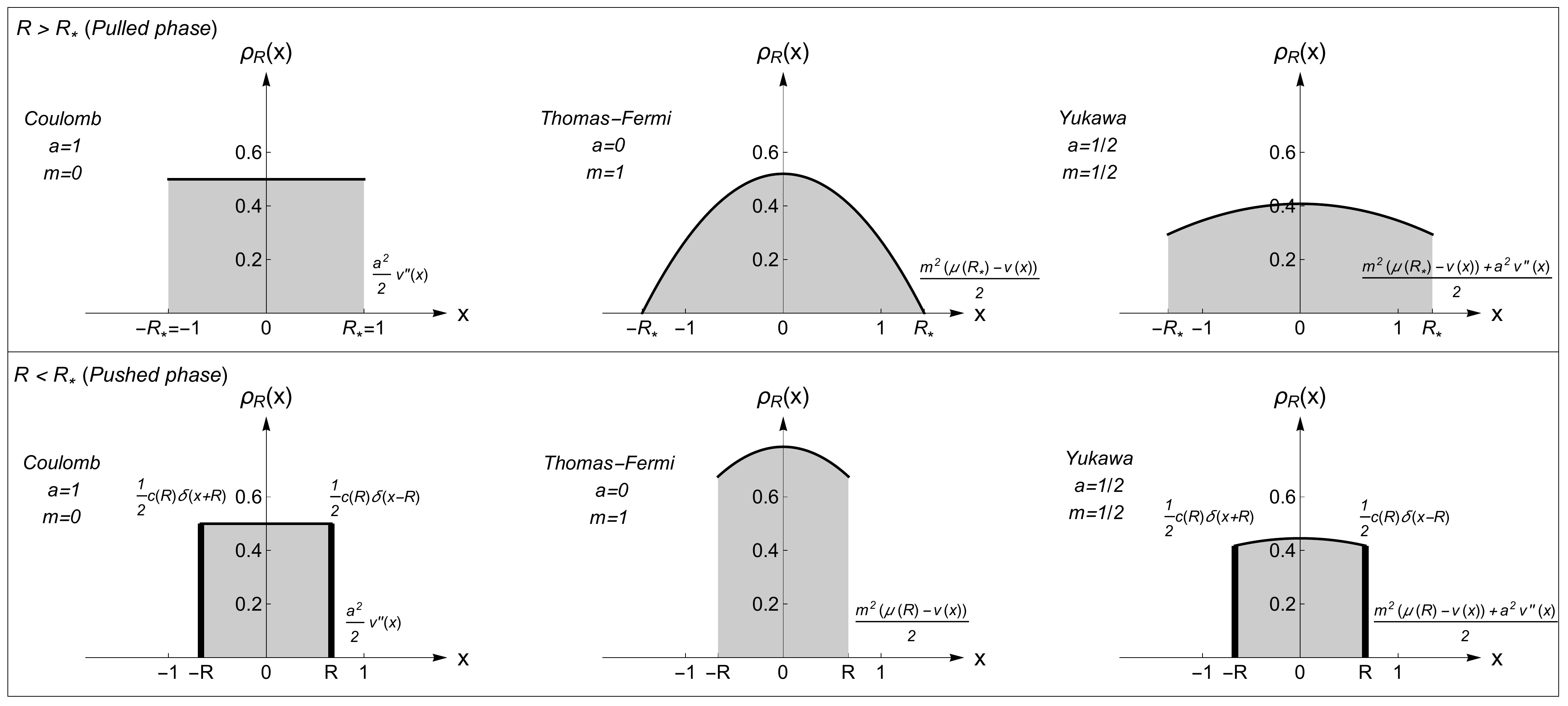}
\caption{Density of a one-dimensional ($d=1$) gas with Coulomb $\varphi(x)=-|x|$, delta $\varphi(x)=2\delta(|x|)$, and Yukawa $\varphi(x)=\exp(-\frac{m|x|}{a})/(am)$ repulsive interactions under quadratic confinement $v(x)=x^2/2$. TOP: the pulled phase ($R_{\star}=1$). BOTTOM: the pushed phase (here $R=2/3$).}
\label{fig:mp}
\end{figure*}

Consider now  the case of a delta-potential $\Phi_d(x)=\Omega_d\delta(x)$ in generic dimension $d\geq1$. (This corresponds to $\mathrm{D}=1$.) The energy~\eqref{eq:EF1} associated to the system is a Thomas-Fermi-like functional 
\be
\EE[\rho]=\frac{\Omega_d}{2}\int(\rho(x))^2\de x+\int V(x)\rho(x)\de x.
\label{eq:energyfunct_d}
\ee
The E-L equations in this case are particularly simple,
and the constrained equilibrium density is
\be
\rho_R(x)=\frac{1}{\Omega_d}\big(\mu(R)-V(x)\big) \mathbbm{1}_{|x|\leq  R\wedge R_{\star}}\ ,
\label{eq:equilibriumTF}
\ee
where $R_\star$---the edge of the support in the pulled phase---is determined by the condition that the gas density~\eqref{eq:equilibriumTF} vanishes at the surface $\rho_{R_{\star}}(R_{\star})=0$, i.e. $R_{\star}$ is the smallest solution of $\mu(R_*)=v(R_*)$. The chemical potential is then fixed by the normalization condition, which yields
\beq
\mu(R)=\frac{d}{(R\wedge R_\star)^d}\left(1+\int_0^{R \wedge R_\star} v(r)r^{d-1}\de r\right).
\eeq 
See Fig.~\ref{fig:mp} for a plot of the gas density in the pulled and pushed phases.
\par
The zero-range nature of the interaction forbids $\delta$-components in the equilibrium measure, \emph{both} in the bulk \emph{and} on the surface (otherwise the energy~\eqref{eq:energyfunct_d} would diverge!). This fact casts serious doubts about the possibility to na\"ively extend the formula~\eqref{eq:rateFgeneral}---derived for the Coulomb gas---to the delta-interaction case, as $c(r)=0$ for the latter. We will come back to this issue later.
\par
After elementary steps, we can write explicitly the excess free energy for the Thomas-Fermi gas as
\be
F(R)=\frac{1}{2} \int_{ R\wedge R_{\star}}^R \big(\mu(r)-v(r)\big)^2r^{d-1}\de r.
\label{eq:LDF_d}
\ee
Quite surprisingly, also in this zero-range model we find that $F(R)$ has a jump in the \emph{third} derivative at $R=R_*$, i.e. $F(R_*)=F'(R_*)=F''(R_*)=0$, while
$ F'''(R\uparrow R_*)=-R_*^{d-1}\left(v'(R_*)\right)^2<0$. Therefore, the critical exponent `$3$' is shared by systems with long-range (Coulomb) and zero-range (delta) interaction. This suggests that the third-order phase transition is even more universal than originally expected.
Moreover, the form of~\eqref{eq:LDF_d}, when compared to~\eqref{eq:rateFgeneral}, strongly suggests that a deeper underlying principle---providing a comprehensive formula for the excess free energy valid for any range of the repulsive potential and in any dimension---should be within reach.
\par
To achieve this goal, we will now turn to the constrained problem for Yukawa (also known as screened Coulomb) interaction~\eqref{eq:Yukawa_eq1vv}, which naturally interpolates between the Coulomb gas and the Thomas-Fermi gas.
\section{Yukawa interaction} Writing equation~\eref{eq:Yukawa_eq1vv} in Fourier coordinates
$\widehat{\Phi}_d(k)=\Omega_d/(a^2|k|^2+m^2)>0,$
one obtains by inversion $\Phi_d(x)=\varphi_d(|x|)$, with 
$
\varphi_d(r)=a^{-2}\Gamma\left(d/2\right)^{-1}\left(m/2ar\right)^{\frac{d}{2}-1}K_{\frac{d}{2}-1}(mr/a)$, where $K_\nu(x)$ is the modified Bessel function of the second kind.
\par
Finding the constrained equilibrium density $\rho_R(x)$ is  a considerable technical challenge, which we have managed to overcome. To the best of our knowledge, the general explicit solution of the variational problem for a constrained gas with Yukawa interaction is new~\footnote{The Yukawa gas in dimensions  $d=3$ in a quadratic potential \emph{without} hard-wall constraint has been investigated in plasma physics~\cite{Henning07}. This special case is recovered by the general solution~\eqref{eq:rhoR_Yukawa}-\eqref{eq:mu_c_Yuk}.}.  
The details of the calculation and its rigorous justification will be published elsewhere~\cite{Cunden_unp}.  
Here we will focus on the physical interpretation and consequences of our results, and will instead provide a derivation of the excess free energy by an energy conservation argument. This in turn will allow us to identify the electrostatic pressure as the order parameter of the phase transition.
\par
Based on the analogy with the Coulomb and Thomas-Fermi gas (the Yukawa interaction being an interpolation between the two extremes), we expect that the equilibrium measure consists of two components: i) a continuous density in the bulk proportional to $(-a^2\Delta+m^2)[\mu(R)-V(x)]$, and ii) a possibly nonzero singular component on the surface in the pushed phase. Hence we write 
\begin{equation}
\rho_R(x)=\!\frac{1}{\Omega_d}\Big[\mathrm{D}\big(\mu(R)-V(x)\big)\mathbbm{1}_{|x|\leq R\wedge R_{\star}}
+\frac{c(R)}{R^{d-1}}\delta(R-|x|) \Big] .
\label{eq:rhoR_Yukawa}
\end{equation} 
\par
The above educated guess can be now proved to be the correct solution by checking that the E-L conditions are satisfied and by uniqueness of the minimiser. 
The chemical potential $\mu(R)$ and the excess charge $c(R)$ are fixed by the normalization of $\rho_R$ and the E-L conditions, which entail performing the integration in~\eqref{eq:E-L} explicitly. Remarkably, the integrals can be evaluated in closed form using properties of the Bessel functions.
The result is that $\mu(R)$ and $c(R)$ are solutions of the following explicit \emph{linear} system
\be
\begin{cases}
& \displaystyle\dfrac{\varphi_d'(R)}{\varphi_d(R)}\mu(R)+\dfrac{c(R)}{a^2R^{d-1}} =\dfrac{\varphi_d'(R)}{\varphi_d(R)}v(R)-v'(R)\\
&\displaystyle\dfrac{m^2R^d}{d}\mu(R)+c(R) =1-a^2\dfrac{v'(R)}{R^{1-d}}+m^2\int\limits_{0}^R\dfrac{v(r)}{r^{1-d}}\de r.
\label{eq:mu_c_Yuk}
\end{cases}
\ee
Evidently, both $\mu(R)$ and $c(R)$ depend on $a$ and $m$. 
\par
In the pulled phase, the singular component on the surface is absent and the equilibrium density $\rho_{R_{\star}}$ is supported in the ball $B_{R_{\star}}$ whose radius is the positive solution of
$c(R_{\star})=0$. In the pushed phase, one can verify that the linear system~\eqref{eq:mu_c_Yuk} has a unique solution $\mu(R)$ and $c(R)$. We conclude that the measure~\eqref{eq:rhoR_Yukawa}, with constants given by~\eqref{eq:mu_c_Yuk}, satisfies the E-L conditions, and is therefore the unique equilibrium configuration of the gas.
The Yukawa equilibrium measure~\eqref{eq:rhoR_Yukawa} interpolates between the Coulomb and delta-interaction cases: i) in the pulled phase, it is discontinuous at the edge and the discontinuity goes to zero as the ratio $m/a$ increases; ii) in the pushed phase, an excess charge condenses on the surface (as in the Coulomb gas) \emph{and} the density in the bulk increases by a constant (as in the Thomas-Fermi gas). See Fig.~\ref{fig:mp} for a comparison of the three cases.
\par
Once $\rho_R(x)$ is known, we can compute its energy $\EE[\rho_R]$. We obtain for the excess free energy the remarkably simple formula
\be
F(R)=\frac{1}{2} \int_{R\wedge R_{\star}}^{R_{\star}} \frac{c(r)^2}{a^2r^{d-1}}\de r.
\label{eq:simple}
\ee
From the above expression, it is yet again straightforward to see that 
$F(R)\sim c'(R_{\star})^2 (R_{\star}-R)^3 / 6 a^2 R_{\star}^{d-1}$, as  $R\uparrow  R_{\star}$.
Moreover,~\eqref{eq:simple} recovers Eq.~\eqref{eq:rateFgeneral} for Coulomb gases (for $m=0$ and $a=1$) and also Eq.~\eqref{eq:LDF_d} for Thomas-Fermi gases---whose excess charge on the surface is zero!---by taking the appropriate limit 
\beq 
\frac{c(r)}{a}\to\left(\mu(r)-v(r)\right)r^{d-1},
\eeq
 as $a\to 0$, with $m=1$.
\par
The universal formula~\eqref{eq:simple}---valid for all dimensions, confining potentials and for any range of the pairwise repulsion---is the main result of this Letter. Can a more physical interpretation of it be found?
\par
\section{Energy conservation and order parameter} From basic principles, the increase in free energy of the constrained gas should match the work $W_{R_{\star}\to R}$ done in a quasi-static compression of the gas (with the system in equilibrium with density $\rho_r(x)$  at each intermediate stage $R\leq r\leq R_{\star}$). In formulae, $F(R)=-W_{R_{\star}\to R}$, with
\be
W_{R_{\star}\to R}=\int_{V_i}^{V_f} \!p\,\de x=\int_{R_\star}^R p(r)\Omega_d r^{d-1}\de r,
\ee
where $V_{i}=\operatorname{vol}(B_{R_{\star}})$ and $V_{f}=\operatorname{vol}(B_{R})$ are the initial and final volumes, and $p(r)$ is the pressure on the gas confined in $B_r$. In other words, $p(r)\Omega_d r^{d-1}\de r$ is the work done on the surface of the ball of radius $r$ being compressed from $r+\de r$ to $r$. 
\par
The pressure is given by the normal force $F_n$ per unit area, $p = \de F_n/\de A$, and $\de F_n$ is in turn equal to the product of the charge contained in a small area $\de A$ on the sphere of radius $r$, times the electrostatic field across $\de A$ (the normal derivative of the potential generated by $\rho_r(x)$). For the amount of charge in $\de A$, this is clearly given by $c(r)\de A/(\Omega_d r^{d-1})$. For the field, one has to integrate the equation $(-a^2\Delta+m^2)\Phi^r =\Omega_d \varrho_r$ over a small cylinder cutting across the surface of the ball. One finds 
that the field is perpendicular to the surface and given by  
$\nabla \Phi^r (x)=c(r) x /a^2 r^{d}$ 
\emph{immediately outside the ball}, while $\nabla \Phi^r=0$ inside the ball (a  consequence of the E-L conditions). The discontinuity of the field across the surface is accounted for by averaging the field inside and outside, which provides an extra factor $1/2$. Putting everything together, we indeed obtain 
\begin{equation}
p(r)=\frac{1}{2\Omega_d} \frac{c(r)^2}{a^2 r^{2d-2}},
\end{equation}
recovering Eq.~\eqref{eq:simple}. 
\par
This basic and universal---albeit previously unnoticed---`energy conservation' argument further elucidates what the appropriate \emph{order parameter} 
of this transition is: the `electrostatic' pressure on the surface of the constrained gas (see Fig.~\ref{fig:transition}).
\par
\section{Conclusions} In summary, the free energies of particle systems with pairwise repulsive interaction of type~\eqref{eq:Yukawa_eq1vv} generically display a third-order singularity across the pulled-to-pushed transition. The order parameter of this phase transition is the  \emph{pressure} of the gas, generated by the surface excess charge. This work considerably broadens the universality of the third-order phase transition, and elucidates the order parameter of the phase transition, leading to the universal formula~\eqref{eq:simple}. 
\par
The findings reported in this Letter raise several questions. For instance, it would be challenging to compute the \emph{subleading corrections} in $N$ to the free energies (a rather standard calculation in random matrix theory~\cite{Alvarez16,Atkin14,Borot11,Cunden16}) and the \emph{crossover scaling functions} (analogues of Tracy-Widom, Gumbel, etc.) between the pulled and the pushed phases. See, e.g.,~\cite{Baxter62,Chafai14b,Dhar17,Ebrahimi18,Lacroix17,Rider03,TW94}. 
\par
It is also worth mentioning that there exists a list of phase transitions (not of pulled-to-pushed type) associated to constrained log-gases in $d=1$ and $d=2$, including the Kazakov-Douglas type~\cite{Douglas93,Forrester11,Levy15}, evaporation~\cite{DePasquale10,Majumdar09,Cunden16c,Cunden16,Texier13}, splitting-merging~\cite{Borot12,Grabsch17,Majumdar09b,Majumdar11,VMB08,Vivo10}, and change of topology~\cite{Allez14,Cunden15b,Cunden16}. An interesting program would be to formulate a unified theory of these phenomena. Further study is in progress.

\ack
The research of FDC is supported by ERC Advanced Grant 669306.
PV acknowledges the stimulating research environment provided by the EPSRC Centre for Doctoral Training in Cross-Disciplinary Approaches to Non-Equilibrium Systems (CANES, EP/L015854/1).
ML was supported by Cohesion and Development Fund 2007--2013 - APQ Research Puglia Region ``Regional program supporting smart specialization and social and environmental sustainability - FutureInResearch.'' 
PF was partially supported by Istituto Nazionale di Fisica Nucleare (INFN) through the project ``QUANTUM.''
FDC, PF and ML were partially supported by the Italian National Group of Mathematical Physics (GNFM-INdAM).
\par

\end{document}